\documentclass[11pt]{article}
\usepackage{amsfonts} 
\usepackage[utf8]{inputenc}
\usepackage{amssymb, amsmath}
\usepackage{tabularx}
\usepackage{multirow}
\usepackage{amsmath} % for 'bmatrix' environment
\usepackage{physics}
\usepackage{cite}

\title{Higgs mode of modified cosmology}

\author{
Metin Arık, Tarik Tok \\
  Department of Physics\\
 Bogazici University\\
   Bebek, Istanbul, Turkey\\
  \texttt{metin.arik@boun.edu.tr}\\
  \texttt{tarik.tok@boun.edu.tr} }

\begin{document}

\maketitle

\begin{abstract}
 
We consider a model where the Standard Model is added to the Einstein Lagrangian together with a Jordan-Brans-Dicke(JBD) coupling. 
 The time-dependent Higgs field has an important role in interpreting the effective gravitational constant, $G_{eff}$. This may lead to two Big Bangs, the first Big Bang characterizes the size of the universe being zero.  At this Big Bang, the value  of the effective gravitational constant is zero and starts decreasing in time through negative values.  During this era, the JBD term is important. In the second Big Bang, the effective gravitational constant passes through infinity to positive values. The negative gravitational constant is interpreted as repulsive gravity. The Lagrangian density provides effective potentials leading to spontaneous symmetry breaking which gives cosmological expectation value of the Higgs field and the Higgs mass which depends on curvature and the Brans Dicke parameter. 
\end{abstract}
\section{Introduction}

Since the Renaissance, two ideas concerning nature of space have dominated our thinking. The first one is that space-time has absolute physical structure with its own nature. This idea disseminated from Descartes' vortex theory\cite{descartes} and Newton's absolute space to aether theories proposing a transmission medium for the propagation of electromagnetic or gravitational forces.
Contrary to this out-fashioned idea, the physical and geometrical properties of empty space are senseless.  The second one is the relativistic point of view in which the only meaningful motion of an object is relative to other objects. After Maxwell's equations and the Michelson-Morley experiment\cite{morley},   Henri Poincare proposed that all interactions  including gravity should be relativistic\cite{bracco}.   In the same year,  Einstein published a paper which established postulates of special relativity\cite{einstein1}. He brought the General Theory of Relativity as the most extraordinary achievement in modern physics 10 years later\cite{eintein2}. In this theory,  gravity is described as distortion of space-time curvature caused by matter or energy.  Moreover,  General Relativity explained the anomalous precession of Mercury's perihelion and also anticipated black holes, gravitational lensing, gravitational waves, and the universe's origin and evolution\cite{wald}. 

General relativity requires both classical and quantum corrections in the strong and weak gravity regimes to explain certain natural events, such as changes in the orbits of binary pulsars, black hole mergers, and the presence of exotic sources\cite{will, Shankaranarayanan}. This evaluation leads to modified gravity theories which may be classified as beyond second order,  extra field contents, non-minimal coupling,  and higher dimensions. In the class of the extra field contents, there is the Jordan-Brans-Dicke theory which is one of the most influential modified gravity theories.

Jordan-Brans-Dicke theory has a scalar–tensor structure which uses both scalar field and tensor field in order to represent gravitational interactions.  Although JBD theory  has some problems related to naturalness\cite{Fujii}, indeterminacy\cite{indeterminacy}, experimental consistency\cite{will} ,and geometric interpretation\cite{frederic}, it is the most famous among modified theories of gravity.  JBD theory besides predicting light deflection and the precession of perihelia\cite{deflection}, may also explain inflation\cite{inflation},  homogeneities\cite{homogen}, dark energy\cite{darkenergy} and stability of stars\cite{stability}. 
In this paper, we consider a hybrid model where the early universe is described by JBD theory and  evolves into an Einsteinian universe with time-independent Newtonian gravitational constant.

In physics, the concept of mass is important as well as space time.  Many areas of physics
can be handled from different perspectives.  The most well-known area where mass is important is particle physics.  The mass in the universe originates from the
masses of fundamental particles.  Until the Higgs boson was discovered by the Atlas\cite{atlas} and CMS\cite{cms} experiments at CERN's Large Hadron Collider,  there had been a contradiction between theoretical framework of particle physics and experimental result of non-zero particle masses, so this theory was required to explain how the particles acquire mass.
According to the Standard Model,  all fundamental particles obtain their masses from the expectation value of the Higgs Field which fills all of space and time.  Moreover,  the Higgs field of the Standard Model could have significantly impacted cosmology, contributing to the Universe's homogeneity, isotropy, and flatness\cite{shaposnikov}. A model in which the Higgs field affects the growth of curvature perturbations and structure of the universe as a spectator discussed by Shaposnikov and others\cite{shaposnikov, Shaposhnikov, nonminimal2, nonminimal3, nonminimal4, nonminimal5, horn} motivates this paper where  we consider a model that hybridizes Einstein's relativity and JBD theory with the Higgs field.

\section{Hybridization of General Relativity and JBD theory with Higgs field}

In modern physics, the most extraordinary achievement which describes geometric properties of gravitation and stages of the universe is the General Theory of Relativity.  On the other hand,  the JBD theory is also consistent in explaining the early stages of the universe. In addition to those theories,  the Higgs field is the only known candidate scalar field to have an important role in cosmology\cite{higgs1, higgs2}.
In this section,  we will discuss how to hybridize general relativity and the JBD theory and assign the Higgs field to be a fundamental field in the formation and evolution of the universe. We will use units where $\hbar=c=1$, and the Planck mass in Einstein's theory is defined by 

\begin{equation}
M_p^2=\frac{1}{8\pi G_N}.
\end{equation}

On the other hand,  in JBD theory, the effective  Newtonian constant of gravitation, $G_{JBD}$, is directly proportional to Brans-Dicke parameter, $\omega$,  and inversely  proportional to square of a time dependent scalar field

\begin{equation}
G_{JBD}
=\frac{\omega}{2\pi  \phi^{2} }
\end{equation} 
where $G_{JBD}$ has the property of effectiveness  because it may change as the universe evolves and dimension of $\phi$ is mass.

We consider a modified theory of gravitation where Einstein's theory is hybridized with 
JBD theory.  In order to describe the effective  Newtonian constant of gravitation, we  suggest a relation

 \begin{equation}
\frac{1}{8\pi G_N}+\frac{1}{4\omega} \phi^{2}
=\frac{1}{8\pi G_{eff}} \label{geff}
\end{equation}
where the effective  Newtonian constant of gravitation, $G_{eff}$, depends on cosmological time\cite{Kaiser}.
We also note that a theory with a negative Brans-Dicke parameter, $\omega<0$,  which we will discuss in section (\ref{sectioneearlyuniverse}) shows that there may be two instants of time in the early universe which can be called two Big Bangs. In the first Big Bang,  the universe starts from zero scale size,  $a=0$, and evolves with a negative gravitational constant, which increases in magnitude and passes to positive values from infinity.
 Note that  for positive $\omega$, as $\phi^2$ increases, $ G_{eff}$ decreases , whereas for negative $\omega$,  as $\phi^2$ decreases $ G_{eff}$ increases.

We know that in the present era,  the value of the Higgs field is much smaller than the Plank mass.  This implies that  the second term in the left hand side of (\ref{geff}) is negligible. This equation may also be interpreted as

 \begin{equation}
M_{peff}^{2}
=\frac{1}{8\pi G_{eff}} \label{peff}
\end{equation}
where $M_{peff}$ is effective Planck mass, which is time-dependent.

Based on these motivations,  the hybridized action is

\begin{equation}\label{lagrangian1}
\mathcal{S}[\Theta, g^{\mu\nu}]= \int\sqrt{g}\left(\frac{1}{2}g^{\mu\nu}\partial_\mu \Theta^\dagger \partial_\nu \Theta-\bigg(\frac{1}{8w} \, \Theta^\dagger\Theta+\frac{M^2_p}{2}\bigg)R \, -\Lambda-V(\Theta)\right)d^4x
\end{equation}
where $g^{\mu\nu}$ is the metric,  $\omega$ is the
dimensionless Brans-Dicke parameter, $M_p$ is the Planck mass, $\Lambda$ is  the cosmological constant,  R is the Ricci scalar, $V(\Theta)$ is the quartic potential

$$V(\Theta)=-\frac{1}{2} m^2 \Theta^\dagger\Theta+\frac{\lambda}{4} (\Theta^\dagger\Theta)^2.$$
and, $\Theta$ is the weak isospin Higgs doublet with four real components

\begin{equation}\label{multiplet}
\Theta=\frac{1}{\sqrt{2}}
\begin{pmatrix}
\phi_1+i\phi_2 \\
\phi_3+i\phi_4
\end{pmatrix}
\end{equation}
where $\phi_a$ corresponds to scalar fields and a=1,2,3,4. This doublet constitutes SO(4) symmetric scalar field space.

One sees that Einstein's theory is obtained in the limit $\Theta\rightarrow0$,  whereas standard JBD theory is obtained when Plank mass is zero.  Regarding  the Higgs multiplet,  $\Theta$, as the fundamental field,  one  notices that the action \ref{lagrangian1},  has the form 

\begin{equation}\label{veff}
\mathcal{S}[\Theta,  g^{\mu\nu}]= \int\sqrt{g}\left(\frac{1}{2}g^{\mu\nu}\partial_\mu \Theta^\dagger \partial_\nu \Theta-V_
{eff}(\Theta, R)\right)d^4x
\end{equation}
where effective potential, $V_
{eff}$, is curvature dependent

\begin{equation}\label{effectivepotential}
V_
{eff}(\Theta, R)=\bigg(\frac{1}{8w} \, \Theta^\dagger\Theta+\frac{M^2_p}{2}\bigg)R \, +\Lambda- \frac{1}{2} m^2 \Theta^\dagger\Theta+\frac{\lambda}{4} (\Theta^\dagger\Theta)^2.
\end{equation}

 We can choose direction of the cosmological expectation value of Higgs field, so three components of the Higgs multiplet, (\ref{multiplet}),  are 

\begin{equation}
\phi_1=\phi_2=\phi_4=0
\end{equation}
as in the Standard Model.  Thus, we identify the scalar field in JBD theory by one chosen real component of the Higgs doublet.

In Minkowski space-time which is static,  the Higgs field,  $\phi$, has to be time independent. However,  in the Friedmann–Lemaître–Robertson–Walker universe which is dynamic,  it is natural to expect that $\phi$ can depend on cosmological time

\begin{equation}
\phi_3=\phi(t).
\end{equation}

The Robertson-Walker metric emphasizes that $\phi$ is necessarily spatially
homogeneous%
\begin{equation}
ds^{2}=dt^{2}-a^{2}\left( t\right) \,\frac{d\vec{x}^{2}}{\left[ 1+\frac {k}{4%
}\vec{x}^{2}\right] ^{2}}  \label{metric}
\end{equation}
\noindent where $k$\ is the curvature parameter with $k=-1$, $0$, $1$\
corresponding to hyperbolic, flat, spherical universes respectively and $a\left(
t\right) $ is the scale factor of the universe.
 The scalar field has dimensions of mass.

For the FLRW metric, one  obtains three equations from the action,  (\ref{lagrangian1}).  Two of them,  (\ref{des}), (\ref{pres}),  are derived by variation with respect to the metric,  $\frac{\delta \mathcal{S}}{\delta g^{\mu\nu}}$, whereas the last can be found by variation with respect to $\phi$,  $\frac{\delta \mathcal{S}}{\delta \phi}$.   After substituting the metric,  the equations\cite{calik} can be written as

\begin{equation}
3\left(\frac{1}{4\omega }\phi ^{2}+M^2_p\right)\,\left( \frac{\dot{a}^{2}}{a^{2}}+\frac{k}{
a^{2}}\right) -\frac{1}{2}\,\dot{\phi}^{2}+\frac{1}{2}\,m^{2}\,\phi ^{2}-\frac{1}{4}\,\lambda\,\phi ^{4}-\Lambda
+\frac{3}{2\omega }\,\frac{\dot{a}}{a}\,\dot{\phi}\,\phi =\rho , \label{des}
\end{equation}%
\begin{equation}
-\left(\frac{1}{4\omega }\phi ^{2}+M^2_p\right)\,\left( 2\frac{\ddot{a}}{a}+\frac{\dot{a}^{2}}{%
a^{2}}+\frac{k}{a^{2}}\right) -\frac{1}{\omega }\,\frac{\dot{a}}{a}\,\dot{%
\phi}\,\phi -\frac{1}{2\omega }\,\ddot{\phi}\,\phi -\left( \frac{1}{2}+\frac{%
1}{2\omega }\right) \,\dot{\phi}^{2}-\frac{1}{2}\,m^{2}\,\phi ^{2}+\frac{1}{4}\,\lambda\,\phi ^{4}+\Lambda=p,
\label{pres}
\end{equation}%
and the equation derived by varying the Higgs field is
\begin{equation}
\ddot{\phi}+3\,\frac{\dot{a}}{a}\,\dot{\phi}-\left[ m^{2}+\frac{3}{2\omega }%
\left( \frac{\ddot{a}}{a}+\frac{\dot{a}^{2}}{a^{2}}+\frac{k}{a^{2}}\right) %
\right] \,\phi +\lambda\phi ^{3}=0\label{fi}
\end{equation}
where $\rho$ is density, $P$ is pressure, and dot denotes $\frac{d}{dt} $.  Scale factor $a$ depends on time so that it can be used as an independent variable .  For the sake of seeing the fate of the scale size of the universe clearly,  one may change the independent variable from time t to scale factor a. We have the relations

\begin{equation}
\dot{\phi}=Ha\phi',\label{var1}
\end{equation}

\begin{equation}
\ddot{\phi}=H^2a^2\phi''+(H'Ha+H^2a)\phi',\label{var2}
\end{equation}

\begin{equation}
\dot{a}=Ha,\label{var3}
\end{equation}

\begin{equation}
\ddot{a}=H'Ha^2+H^2a,\label{var4}
\end{equation}
where $H$ is Hubble parameter depends on scale factor, a,  and prime denotes $\frac{d}{da}$.  After substituting (\ref{var1}), 
(\ref{var2}), (\ref{var3}) and (\ref{var4}) into (\ref{des}), (\ref{pres}), and (\ref{fi}),  they turn into

\begin{equation}
3\left(\frac{1}{4\omega }\phi ^{2}+M^2_p\right)\,\left( H^2+\frac{k}{
a^{2}}\right) -\frac{1}{2}\,H^2a^2\phi'^{2}+\frac{1}{2}\,m^{2}\,\phi ^{2}-\frac{1}{4}\,\lambda\,\phi ^{4}-\Lambda
+\frac{3}{2\omega }\,H^2\,a\phi'\phi =\rho , \label{dess}
\end{equation}%
\begin{equation}
\begin{split}
-\left(\frac{1}{4\omega }\phi^2 +M^2_p\right)\left( H'Ha+3H^2+\frac{k}{a^{2}}\right) -\frac{1}{2\omega }\,\left( H'Ha^2+3H^2\,a \right)
\phi'\phi -\frac{1}{2\omega }H^2a^2\phi''\phi \\
\\
-\left( \frac{1}{2}+\frac{%
1}{2\omega }\right) H^2a^2\phi'^{2}-\frac{1}{2}m^{2}\phi ^{2}+\frac{1}{4}\,\lambda\,\phi ^{4}+\Lambda=p,
\label{press}
\end{split}
\end{equation}%
\begin{equation}
H^2a^2\phi''+\left[H'Ha^2+4H^2 a\right]\phi' -\left[ m^{2}+\frac{3}{2\omega }%
\left( H'Ha+2H^2+\frac{k}{a^2}\right) %
\right] \,\phi +\lambda\phi ^{3}=0 ,\label{fii}
\end{equation}
respectively.

\section{Dynamics of the Universe}\label{sectioneearlyuniverse}
In order to understand the dynamics of the universe, we have to solve the gravitational field equation

\begin{equation}  \label{gravitational}
G^{\mu}_{\nu}= 8\pi G_{eff} T^{\mu}_{\nu}.
\end{equation}
$(0,0)$ component of this which may be called the modified Friedmann-Lemaitre equation can be written as

\begin{equation}
H^2=( G_N+\frac{\omega}{2\pi \phi^2
}) \rho+\frac{\Lambda}{3}-\frac{k}{a^2}
\end{equation}
where, energy density of the universe, $\rho$, is the key parameter to understand the dominance of the various kinds of components of the universe. It may be expanded as

\begin{equation}\label{rhoexpand}
 \rho=\frac{R_{\rho}}{a^4}+\frac{M_{\rho}}{a^3}+\frac{S_{\rho}}{a^2}+\frac{W_{\rho}}{a}+{\Lambda}_{\rho} 
\end{equation}
where $R_{\rho}$, $M_{\rho}$,  $S_{\rho}$,  $W_{\rho}$,  $\Lambda_{\rho}$ are dimensional constants for radiation, matter, cosmic strings, cosmic walls and cosmological constant dominated cases respectively.

Notice that (\ref{dess}), (\ref{press}), (\ref{fii}) are complicated. Therefore, we will try to solve them perturbatively by considering anzatzes.

\subsection{Cosmological Solution}
\label{Cosmological Solution}

Equations (\ref{dess}), (\ref{press}) and (\ref{fii}) contain the Hubble parameter which has crucial properties of the universe including its expansion, evolution, and structure, as well as linking cosmological observations to theoretical models. It's dimension is $\frac{1}{length}$, then the simplest anzatz that one can formulate which is compatible for both the early universe and the late universe is

\begin{equation}\label{hubblee}
H^2=\frac{h_1}{a^2}+\frac{h_2}{a}+h_3
\end{equation}
where $h_1$, $h_2$, and $h_3$ are constants. $h_1$ can be interpreted as the velocity of the expanding early universe.

In the Standard Model, the solution of the potential for constant scalar field is compatible only for Minkowski spacetime. However, if we add the Standard Model to General Relativity, it would be natural to write a Brans-Dicke term to the action due to the scalar field. When we look at the solution, we obtain constant  cosmological density parameter, (\ref{dess}) . We see that it cannot be possible for the constant scalar field, because radiation, $a^{-4}$, and matter, $a^{-3}$, terms are then zero. In addition to this, if the scalar field goes as $a^1$, its density parameter, (\ref{dess}), has zero matter, $a^{-3}$, term.  Therefore, we propose

\begin{equation}\label{fieldd}
\phi=\frac{f_1}{a}+f_2
\end{equation}
where $f_1$, and $f_2$ are constants. This is the simplest behaviour of the scalar field, $\phi$, which could be consistent with our model.

Anzatzes (\ref{hubblee}) and (\ref{fieldd}) can be substituted in the field equation(\ref{fii}), which can be expanded in the inverse  power of {$a$}

\begin{equation}
0=\frac{A_{L}}{a^3}+\frac{B_{L}}{a^2}+\frac{C_{L}}{a}+D_{L}
\label{fieldexplicit}
\end{equation}

where

\begin{equation}
A_{L}=f_{1}\big[-h_1\big(\frac{3}{2\omega}+1\big)-\frac{3k}{2\omega}+\lambda f^2_{1}\big]
\end{equation}

\begin{equation}
B_{L}=-f_1\big[\frac{3}{2}h_2\big(\frac{3}{2\omega}+1\big)\big]-f_2\big[\frac{3}{2\omega}(h_1+k)-3\lambda f^2_{1}\big]
\end{equation}

\begin{equation}
C_{L}=-f_1\big[2(\frac{3}{2\omega}+1)h_3-3\lambda f^2_{2}+m^2\big]-f_2\frac{9}{4\omega}h_2
\end{equation}

\begin{equation}
D_{L}=-f_2\big[\frac{3}{\omega}h_3-\lambda f^2_{2}+m^2\big].
\end{equation}

In a similar manner, coefficients of  energy density(\ref{rhoexpand}) can also be expressed as

\begin{equation}
R_{\rho}=f^2_{1}\big[-h_1\big(\frac{3}{4\omega}+\frac{1}{2}\big)+\frac{3k}{4\omega}-\frac{\lambda}{4}f^2_{1}\big]
\end{equation}

\begin{equation}
M_{\rho}=-f^2_{1}h_2\big[\frac{3}{4\omega}+\frac{1}{2}\big]+f_{1}f_{2}\big[\frac{3k}{2\omega}-\lambda f^2_{1}\big]
\end{equation}

\begin{equation}
W_{\rho}=f^2_{1}\big[-h_3\big(\frac{3}{4\omega}+\frac{1}{2}\big)+\frac{m^2}{2}\big]+f^2_{2}\big[\frac{3}{4\omega}(h_1+k)-\frac{3}{2}\lambda f^2_{1}\big]+3M^{2}_p(h_1+k)
\end{equation}

\begin{equation}
S_{\rho}=f_{1}f_{2}m^2+f^2_{2}  \big[\frac{3}{4\omega}h_2-\lambda f_{1}f_{2}\big]+3M^{2}_p h_2
\end{equation}

\begin{equation}
\Lambda_{\rho}=f^2_{2}  \big[\frac{3}{4\omega}h_3+\frac{m^2}{2}-\frac{\lambda}{4}f^2_{2}\big]+3M^{2}_p h_2+\Lambda.
\end{equation}

We note that the Weak Energy Condition(WEC) which is a crucial concept in General Relativity implies that the energy density has to be non-negative,  $\rho\geq 0$. We also assume that the partial energy densities\cite{partialenergy} are also non-negative. Therefore, coefficients in (\ref{rhoexpand}) have to be non-negative.

In the early universe, the Robertson–Walker scale factor, $a(t)$, is so small that higher order terms in the field equation(\ref{fieldexplicit}) and energy density(\ref{rhoexpand})  become negligible. Under this condition, they
turn into 

\begin{equation}
0\approx\frac{f_{1}\big[-h_1\big(\frac{3}{2\omega}+1\big)-\frac{3k}{2\omega}+\lambda f^2_{1}\big]}{a^3}+\frac{-f_1\big[\frac{3}{2}h_2\big(\frac{3}{2\omega}+1\big)\big]-f_2\big[\frac{3}{2\omega}(h_1+k)-3\lambda f^2_{1}\big]}{a^2}
\end{equation}

\begin{equation}
 \rho\approx\frac{f^2_{1}\big[-h_1\big(\frac{3}{4\omega}+\frac{1}{2}\big)+\frac{3k}{4\omega}-\frac{\lambda}{4}f^2_{1}\big]}{a^4}+\frac{-f^2_{1}h_2\big[\frac{3}{4\omega}+\frac{1}{2}\big]+f_{1}f_{2}\big[\frac{3k}{2\omega}-\lambda f^2_{1}\big]}{a^3}.
\end{equation}

We have found a positive Brans-Dicke parameter for the closed FLRW universe. In contrast, a negative Brans-Dicke parameter has been obtained for the open FLRW universe, with  $\omega= -3/2$ indicating the presence of a singularity.  However, no solution that satisfies WEC exists for the flat universe within this framework.
Note that when the Higgs field gets the value 

\begin{equation}\label{phi2}
 \phi^2 \rightarrow - \frac{\omega M_p^2}{4},
\end{equation} the effective Newtonian constant of gravitation , $G_{eff}$,  goes to infinity. This situation may be called the second Big Bang.

In the Standard Model of particle physics, which is formulated in Minkowski spacetime, the value of the quartic coupling constant, $\lambda$, has been measured to be approximately 0.13\cite{datagroup}. Because of this value, $f_2$ in (\ref{fieldd}) becomes imaginary and this is incompatible in our model when we try to solve equation (\ref{fieldexplicit}) as a whole such that energy density (\ref{rhoexpand}) is partially non-negative. The condition for the field, $\phi$ to be real could be satisfied only if $\lambda$ is large which indicates an unstable universe \cite{unstablecoupling}. On the other hand, the exact solution can be obtained for negative partial energy densities\cite{negative1, negative2, negative3, negative4, negative5} or for scalar field dependent Brans Dicke parameter\cite{timedep1,timedep2}.
Moreover, this solution indicates nonzero string and cosmic wall contributions to the expanding universe. For late universe, standard cosmology results are obtained as we expected.

 \hfill \break

\subsection{Higgs mass}
The curvature dependent effective potential can be written in terms of the Higgs field 

\begin{equation}\label{effectivepotential}
V_
{eff}(\phi, R)=\bigg(\frac{1}{8w} \, \phi^2+\frac{M^2_p}{2}\bigg)R \, +\Lambda- \frac{1}{2} m^2 \phi^2+\frac{\lambda}{4}\phi^4.
\end{equation}

The minimum of the effective potential is  given by

\begin{equation}\label{minn}
\upsilon=\frac{1}{\sqrt{\lambda}}\big(m^2-\frac{R}{4\omega}\big)^{\frac{1}{2}}.
\end{equation}
After performing a perturbative calculation, the Higgs mass is given by

\begin{equation}
m_{Higgs}=\big(4 m^2-\frac{R}{\omega}\big)^{\frac{1}{2}}.
\end{equation}

Note that Higgs mass depends on the Ricci scalar and the Brans-Dicke parameter. At present time, it's value should be compatible with the Standard Model.

\section{ Conclusion}
We propose a novel cosmological model hybridizing Einstein’s relativity and Brans-Dicke theory with the Higgs scalar sector. Variational equations have been obtained. They are too complicated to solve exactly. In order to get an idea, we propose the simplest anzatzes satisfying presence of radiation and matter in early era of the universe. We have found that positive and negative Brans-Dicke parameter values are consistent with our model. In the early universe, negative values of Brans Dicke parameter implies two Big Bangs. For the late era, this model turns into standard cosmology. By taking the minimum of the effective potential, the Higgs mass’ dependence on Ricci scalar and Brans Dicke parameter  have also been  calculated.

\bibliography{references}

\end{document}